\documentstyle[11pt]{article}

\newcommand{\be}{\begin{equation}}
\newcommand{\ee}{\end{equation}}
\newcommand{\ba}{\begin{array}}
\newcommand{\ea}{\end{array}}
\newcommand{\bc}{\begin{center}}
\newcommand{\ec}{\end{center}}
\newcommand{\bi}{\begin{itemize}}
\newcommand{\ei}{\end{itemize}}

\newcommand{\disregard}[1]{{}}

\def\bild#1\over#2{\mathrel{\mathop{\kern0pt #1}\limits_{#2}}}

\begin{document}
\centerline {{\bf DENSITY CORRELATIONS OF MAGNETIC IMPURITIES AND DISORDER  \rm}}
\vskip 2cm                 
\centerline{{\bf Jean DESBOIS, Cyril FURTLEHNER\rm } and    
{\bf St\a'ephane OUVRY \rm }\footnote{\it  and
LPTPE, Tour 12, Universit\'e Paris  6 / electronic e-mail: 
OUVRY@FRCPN11}}
\vskip 1cm
\centerline{{Division de Physique Th\'eorique \footnote{\it Unit\a'e de Recherche  des
Universit\a'es Paris 11 et Paris 6 associ\a'ee au CNRS},  IPN,
  Orsay Fr-91406}}
\vskip 3cm
{\bf Abstract : We  consider an  electron coupled to 
a random distribution of point 
vortices in the plane (magnetic impurities). We analyze the effect of 
the magnetic impurities on the density of states
of the test particle, when the magnetic impurities have a 
spatial probability distribution governed by Bose or Fermi 
statistic at a given temperature.
Comparison is made with the Poisson distribution, showing that the zero 
temperature Fermi distribution
corresponds to less disorder. A phase diagram describing   isolated
impurities versus  Landau level oscillations is proposed.}

\vskip 1cm

IPNO/TH 96-03

June  1996

\vfill\eject

{\bf 1. Introduction}

\hfill\break
The problem of a 2 dimensional electron gas coupled to
a static random magnetic field has been a subject of 
interest in the past few years [1]. Particular attention has
been paid to localisation properties of such systems. In the case of
Gaussian disorder with zero mean,  all states seem 
to be localised [2].
On the contrary, they are delocalised in the case of an uniform
magnetic field. Therefore, 
the question arises about the role played by a mean-field 
description of a random
magnetic field [3,4]. 

The model we are interested in  consists of a planar electron of electric
charge $e$
coupled to  a random magnetic field \cite{mdis}. 
What we mean by random magnetic field
is a distribution
of infinitely thin vortices carrying a flux
$\phi$, modelizing some sort of magnetic impurities, characterized  by
the dimensionless Aharonov-Bohm coupling
$\alpha=e\phi/2\pi$ (i.e. $\phi$ in unit of the quantum of flux).
This system is periodic in $\alpha$ with period $1$
and since there is no privileged orientation of the plane,
it is invariant by changing $\alpha$
into $-\alpha$, implying that $\alpha$ can be restricted
to the interval $[0,1/2]$.

 In order to study the effect of statistics on the disordered
magnetic impurity systems, one may  evaluate perturbatively 
in $\alpha$ the average one
electron partition
function,  i.e the Laplace transform of the average density of states.

In previous works [4],
we focused on magnetic impurities obeying
a Poisson distribution (which actually corresponds to the Bose case
at zero temperature, or at infinite temperature). We observed in particular a
transition for $\alpha_c\simeq 0.35$ between an
almost free density of states for an isolated impurities system
($\alpha>0.35$) and an 
oscillating  Landau like density of states ($\alpha<0.35$).

In this letter, 
we will consider  the random magnetic impurities system as a gas of particles
of a given density $\rho$, 
with a distribution obeying Fermi or Bose statistics at a temperature
$T_v$.
We will first properly define 
the perturbative expansion of the average partition function. Then, we
will explicitly compute at
order
$\alpha^2$  contributions to the average partition function. 
We will argue that in the Fermi
case, at zero temperature,  the average density of states 
 always displays 
Landau like oscillations,
implying that there is 
no transition between an isolated impurity disordered  phase and a
mean magnetic field phase.
This result is  not unexpected, since Fermi statistics
clearly tends to homogeneize the impurity configurations, leading to a less
disordered situation.
\hfill\break
\hfill\break   
{\bf 2. The Model} 

{\it 2.1. General Formalism}

Let us consider an electron  coupled to a random magnetic field
given by a distribution $\rho({\bf r})$ of magnetic
impurities. This means that  $\rho({\bf r})d {\bf r}$ is the number of
impurities at position $\bf r$ in the infinitesimal volume $d^2\bf r$.
The  Hamiltonian is given by
\be H={1\over 2m}\left({\bf p} - \alpha\int d^2{\bf r}'\rho({\bf r}')
{{\bf k}\times({\bf r}-{\bf r}')
\over\vert{\bf r} -{\bf r}'\vert^2}\right)^2\pm{\alpha\over m}
\rho({\bf r})\ee
where we have explicitely taken into account the coupling of the magnetic
field to the spin-up $(+)$ or 
down $(-)$ degree of freedom of the electron
endowed with a magnetic moment $\mu=-{e\over 2m}$ (thus an electron
with a gyromagnetic factor g=2).

In the case of a discrete distribution
$\rho({\bf r})=\sum_i \delta({\bf r}-{\bf r}_i)$,     where
the index $i$ indices the impurities,  the 
spin-term is a sum of contact terms. It
corresponds to a choice of a peculiar self adjoint extension [5]:
in the $(+)$ case, the wave functions vanish at the location of the
impurities (hard-core boundary condition), whereas in the $(-)$ case
 singular wave functions are considered at the location of the impurities
(attractive-core conditions). In order to extract the short distance
behaviour of the wave functions, a non-unitary wave
function redefinition has been used [6]:
\be \psi^{\pm}_N({\bf r})=\prod_{i=1}^N\vert{\bf r}-{\bf r}_i\vert^{\pm\alpha}
\tilde\psi^{\pm}_N({\bf r})\ee

The generalisation of this transformation to a continuous distribution
$\rho(\bf r)$ is
\be \psi^{\pm}({\bf r})=e^{\pm\alpha\int d^2{\bf r}'\rho({\bf r}')
\ln\vert{\bf r}-{\bf r}'\vert}
\tilde\psi^{\pm}({\bf r})\ee
The Hamiltonian $\tilde H^{\pm}$ acting on $\tilde\psi^{\pm}(\bf r)$
rewrites 
\be \tilde H^+=-{2\over m}\partial_z\partial_{\bar z} -{2\alpha\over m}\int
dz'd\bar z'{\rho(z',\bar z')\over\bar z-\bar z'}\partial_z\ee
\be \tilde H^-=-{2\over m}\partial_z\partial_{\bar z} +{2\alpha\over m}\int
dz'd\bar z'{\rho(z',\bar z')\over z-z'}\partial_{\bar z}\ee
where the complex coordinates in the plane have been used
$z=x+iy$, $\partial_z={1\over2}(\partial_x-i\partial_y)$
and $dzd\bar z=d^2{\bf r}$.
$H$ or $\tilde H$ can be used indifferently to compute the partition
function, since it is by definition the trace of a function of $H$.
In the sequel we will concentrate on the spin up coupling, keeping in mind
that the spin down analysis could be easily done following the same lines.

Up to now $\rho(\bf  r)$ has not been  specified. If a Poissonian
distribution is choosen,  $\rho(\bf r)$
is defined by its cumulents 
\be \overline{\rho({\bf r}_1)...\rho({\bf r}_k)}=\rho\delta({\bf r}_1-{\bf r}_2)
\delta({\bf r}_2-{\bf r}_3)...\delta({\bf r}_{k-1}-{\bf r}_k)\ee
Here however, we deal with quantum statistics for
the impurities themselves, so $\rho(\bf  r)$ has to be defined as
\be \rho(\bf r)=\psi^+(\bf r)\psi(\bf r)\ee
with
\be \psi({\bf r})={1\over 2\pi}\int d^2{\bf k} a({\bf k})e^{-i{\bf kr}}\ee
\be \psi^+({\bf r})={1\over 2\pi}\int d^2{\bf k} a^+({\bf k})e^{i{\bf kr}}\ee
$a^+(\bf k)$ and $a(\bf k)$ are the creation and annihilation 
Fock space particle operators , with the commutation rules 
\be [a({\bf k}),a^+({\bf k}')]=\delta({\bf k}-{\bf k}')\ee
for bosons, and
\be \{a({\bf k}),a^+({\bf k}')\}=\delta({\bf k}-{\bf k}')\ee
for fermions.\hfill\break
Thus, the impurities, considered as a quantum gas, have a temperature
$T_v$ and a chemical potential $\mu$, which determines their mean 
density $\rho$. 
 The average over disorder
of an operator $Q$ consists in 
\be <Q>={T_r[e^{-\beta_vH_v}Q]\over T_r[e^{-\beta_vH_v}]}\ee
where the impurity second quantized Hamiltonian 
\be H_v=\int d^2{\bf k}({k^2\over 2m}-\mu)a^+({\bf k})a({\bf k})\ee
describes the equilibrium of the impurity gas in the grand-canonical
ensemble.
Note that we consider here quenched impurities, which
are in thermodynamical equilibrium.
Note also that the Poissonian distribution $dP({\bf r}_i)=d{\bf r}_i/V$
can be seen as
the particular case of Bose distribution at  $T_v=0$. The impurities
indeed condensate in the zero energy $N$-body wave function $\psi({\bf
r}_1,{\bf r}_2,\ldots ,{\bf r}_N)=({1\over\sqrt V})^N$,
leading to the $N$-impurity Poissonian  distribution
$dP= \psi^*({\bf
r}_1,{\bf r}_2,\ldots ,{\bf r}_N) \psi({\bf
r}_1,{\bf r}_2,\ldots ,{\bf r}_N) d{\bf r}_1d{\bf r}_2\ldots d{\bf r}_N$.
\hfill\break

We wish to evaluate perturbatively the average one electron
partition function at inverse temperature $\beta$
$$ <{\bf r}\vert e^{-\beta\tilde H}\vert{\bf r}>=
\sum_{p=0}^\infty ({2\alpha\over m})^p\int_0^\beta d\beta_1...
\int_0^{\beta_{p-1}} d\beta_pG_{\beta-\beta_1}({\bf r},{\bf r}_1)
\psi^+({\bf r'}_1)\psi({\bf r'}_1)$$
\be\label{devel} {\partial_{z_1}\over\bar z_1-\bar z'_1}
G_{\beta_1-\beta_2}({\bf r}_1,{\bf r}_2)...
\psi^+({\bf r'}_p)\psi({\bf r'}_p){\partial_{z_p}\over\bar z_p-\bar z'_p}
G_{\beta_p}({\bf r}_p,{\bf r})\ee
where integrations over the position variables  ${\bf r}_i$ and ${\bf r'}_i$
are implicit.  $G_\beta({\bf r},{\bf r'})$ is the free electron propagator
\be G_\beta({\bf r_1},{\bf r_2})={m\over 2\pi\beta}
e^{-{m\over 2\beta}\vert{\bf r_1}-{\bf r_2}\vert^2}\ee
Averaging over disorder yields expressions like
$$<\rho({\bf r}'_1)...\rho({\bf r}'_p)>=
{Tr[e^{-\beta_v H_v}\psi^+({\bf r'}_1)\psi({\bf r'}_1)...
\psi^+({\bf r'}_p)\psi({\bf r'}_p)]\over Tr[e^{-\beta_v H_v}]}$$
which can be evaluated using the contractions
\be \pm g_f^{\pm}({\bf r},{\bf r'})=<\psi({\bf r})\psi^+({\bf r}')>=
\int {d^2{\bf k}\over 4\pi^2}(1\pm n_{\bf k}) e^{i\bf k({\bf r-r}')}\ee    
\be g_b^{\pm}({\bf r},{\bf r}')=<\psi^+({\bf r})\psi({\bf r}')>=
\int {d^2{\bf k}\over 4\pi^2} n_{\bf k} e^{i{\bf k}({\bf r-r}')}\ee
$n_{\bf k}$ stands for the Bose-Einstein (upper sign) or Fermi-Dirac
(lower sign) 
distributions. 
\be n_{\bf k}={1\over e^{\beta_v(k^2/2m-\mu)}\mp 1}\ee
One has the relation
\be g_f^{\pm}({\bf r},{\bf r'})=
g_b^{\pm}({\bf r},{\bf r'})\pm\delta({\bf r}-{\bf r}')\ee

To summarize, the perturbative  expansion of the one electron average partition function
can be 
represented in terms of Feynmann diagrams given by rules quite analogous to
those of finite temperature second quantised formalism: \hfill\break
electron line:{\hskip 4cm}$G_\beta({\bf r}_i,{\bf r}_j)$\hfill\break
forward impurity line:{\hskip 3cm}$g_f^{\pm}({\bf r}'_i,{\bf r}'_{\sigma(i)})$
\hfill\break
backward impurity line:{\hskip 2.5cm}$g_b^{\pm}({\bf r}'_i,{\bf r}'_{\sigma(i)})$
\hfill\break
impurity loop:{\hskip 4cm}$g_b^{\pm}({\bf r}'_i,{\bf r}'_i)=\rho$\hfill\break
electron-impurity vertex:{\hskip 2.5cm}${2\alpha\over m}{1\over\bar z_i-\bar z'_i}\partial_{z_i}$
\hfill\break
For a given diagram of order $p$, the electron propagates from
 its initial to its final position $\bf  r$ via ${\bf  r}_p$,
 ${\bf r}_{p-1}$,...${\bf  r}_1$, the location of
electron-impurity interaction, with temperatures $0$, $\beta_p$,...,
$\beta_1$, $\beta$. The $p$ vortices located at position ${\bf r}'_1$... 
${\bf r}'_p$,
undergo a permutation $\sigma$. At each $\sigma(i)$, 
it corresponds a vortex line:

backward line if $\sigma(i)>i$

forward line if $\sigma(i)<i$

and a loop if $\sigma(i)=i$.

In the Fermi case, each diagram is affected by
the signature of the permutation $\sigma$.\hfill\break
The dimensionless parameters at work are 
the rescaled average density $\lambda^2\rho$ in unit of the electron thermal
wavelength $\lambda^2=2\pi\beta/m$, the rescaled average density
$\lambda_v^2\rho$ in unit of the impurity thermal
wavelength $\lambda_v^2=2\pi\beta_v/m$, and  the Aharonov-Bohm coupling
constant $\alpha$. 

Clearly, one expects that in the limit $\beta_v\to
0$, i.e. Boltzmann statistics, with uncorrelated randomly dropped impurities,
one recovers Poisson distribution. Also, as already advocated above,
one expects that in the limit $\beta_v\to \infty$, in the Bose case, one has
again the Poisson distribution, whereas  the Fermi distribution
leads to a less disordered situation.
In the sequel, one will concentrate on the relative interplay between the
dimensionless parameters $\lambda^2\rho$, $\lambda^2_v\rho$,
$\alpha$ to study the phase diagram of the magnetic
impurity system.

\hfill\break
{\it 2.2. Mean-field expansion}
\hfill\break
\hfill\break
Consider first diagrams that are entirely built by 
impurity loops (Fig. 1a). These diagrams do not involve 
the many-body statistical correlations of the
impurity distribution, and are thus independant
of the statistic. Therefore, they  yield the same contribution as in 
the Poisson case [4].\hfill\break
The  order $\alpha^p (p>0)$ term writes
$$-{1\over \lambda^2}{\zeta(1-p)\over(p-1)!}
(-\lambda^2\rho\alpha)^p$$
Summation over $p$ yields as it should
the partition function per unit volume of the mean magnetic field
$e<B>=2\pi\rho\alpha$
(i.e. in the mean-field limit
$\alpha\to 0, \rho\to\infty, \rho\alpha$ finite)
\be \label{mean}Z_{<B>}={e<B>\over2\pi}{1\over 2\sinh{\beta}{e<B>\over2 m}}
\exp(-{\beta}{e<B>\over 2 m})\ee 
The global positive 
shift in the Landau spectrum is a direct manifestation of the
hard-core boundary conditions on the wavefunctions \cite{mdis}.
\hfill\break
\hfill\break
{\it 2.3. Second order expansion}
\hfill\break
\hfill\break
Non trivial diagrams (Fig.1b),  i.e. not  mean-field diagrams, appear at second
order in $\alpha$.
Let us denote the fugacity by $z=\mp e^{\beta_v\mu}$.
In the interval $z\in ]-1,1]$ one has the expansion
\be n({\bf k})=\pm\sum_{n=1}^\infty (-z)^n e^{-n\beta_v{k^2\over 2m}}\ee
As a result
\be \label{bobo}g_f({\bf r},{\bf r}')=\delta({\bf r}-{\bf r}')+\sum_{n=1}^\infty
(-z)^n G_{n\beta_v}({\bf r},{\bf r}')\ee   
\be \label{bobo'}g_b({\bf r},{\bf r}')=\pm\sum_{n=1}^\infty
(-z)^n G_{n\beta_v}({\bf r},{\bf r}')\ee
The density $\rho$ is related to $z$ by
\be \rho=\mp{1\over \lambda_v^2}\ln (1+z)\ee
Using (\ref{bobo},\ref{bobo'}), the diagram of Fig. 1b has the contribution
\be \label{diag} D(z)={\rho\alpha^2\over2}-{m\alpha^2\over 4\beta_v}
\sum_{n=1\atop p=1}^\infty
{(-z)^{n+p}\over n+p}\left(1-\int_0^1 dx{x_{np}\over x_{np}+x(1-x)}\right)\ee
where $x_{np}={np\over n+p}{{\beta_v\over \beta}}$.\hfill\break

At high temperature ($T_v>T$), in the Boltzmann limit, i.e. $z\to 0$ 
both in the Bose
and Fermi cases, one gets $D(z)\to \rho\alpha^2/2$, 
which is precisely the Poisson distribution result.
The  correction reads
\be D(z)={\rho\alpha^2\over 2}(1-{\pi\over2}\lambda_v^2\rho)
+\ldots \ee

At low temperature   ($T_v<T$),
$D(z)$ rewrites
\be\label{diag'} D(z)={\rho\alpha^2\over 2}
\left[1\mp\sum_{q=1}^\infty\sum_{m=0}^{q-1}
(-{\lambda^2\rho})^q{(q!)^2\over (2q+1)!}C_{q-1}^m
{h_{q-m}(z)h_{m+1}(z)\over [\ln (1+z)]^{q+1}}\right]\ee
where the functions $h_q(z)$'s are defined in the interval $]-1,1]$ by
\be\label{toto} h_q(z)=\sum_{n=1}^\infty{(-z)^n\over n^q}\ee
The low temperature limit corresponds to  $z\to+\infty$ 
in the Fermi case
and $z\to-1$ in the Bose case. 
It happens that (\ref{toto}) can be analytically continuated in 
$]-1,+\infty[$ by noticing that
\be h_0(z)=-{z\over 1+z}\ee
and by using the recursive relation 
\be h_{q+1}(z)=\int_0^z{h_q(x)\over x} dx\ee
As a result, $D(z)$ in (\ref{diag'}) is valid in the interval 
$z\in ]-1,+\infty[$.
It follows that in the Fermi case and  in
the low temperature limit $T_v\to 0$, one can use the expansion
\be \label{h} h_q(z)=-{(\ln z)^q\over q!}+2\sum_{n=1}^{E(q/2)}h_{2n}(1)
{(\ln z)^{q-2n}\over(q-2n)!}-(-1)^q\sum_{n=1}^{\infty}{(-1)^n\over n^q}
{1\over z^n}\ee
which is valid for $z>1$.

Which contributions does $D(z)$ yield in the low $T_v$ limit? In the Bose
case, some care is required, since in (\ref{diag'}) the limit $z\to 0$
 cannot be interchanged with summations. We checked numerically
that (\ref{diag'})
indeed yields $\rho\alpha^2/2$, i.e. the Poissonian result as expected. 
One obtains for the average
partition function  $Z$ at
order $\alpha^2$
\be\label{bibi} Z={1\over\lambda^2}(1-{1\over2}\lambda^2\rho\alpha
+{1\over2}({\lambda^2\rho})^2\alpha^2
({1\over6}+{1\over\lambda^2\rho})+\ldots )\ee

In the Fermi case, using the identity
$ \sum_{i=1}^{q-1}C_{m=0}^{q-1}{1\over (q-m)!}{1\over (m+1)!}={2q!\over
(q+1)!(q!)^2}$, (\ref{diag'},\ref{h}) yields 
\be\label{D2} \lim_{z\to+\infty}D(z)={\rho\alpha^2\over2}[1-{1\over\lambda^2\rho}
(1+\lambda^2\rho-e^{-\lambda^2\rho}-2\sqrt{\lambda^2\rho}
\int_0^{\sqrt{\lambda^2\rho}} dy e^{-y^2})\ee
At this point one can consider either a low impurity density limit
$\lambda^2\rho<< 1$, or a high impurity density limit  $\lambda^2\rho>> 1$.
At low density
$\lambda^2\rho<< 1$, one finds that
the average
partition function per unit volume rewrites
as \be\label{Zlow}
Z={1\over\lambda^2}(1-{1\over2}\lambda^2\rho\alpha+
{1\over2}(\lambda^2\rho)^2\alpha^2({1\over\lambda^2\rho}+{1\over30}\lambda^2
\rho)+\ldots )\ee
At  order $\alpha^2$, the $\rho^2\alpha^2$ term
is missing,  a situation quite different from the Poissonian case
(\ref{bibi}),
where this mean-field term precisely 
 dominates in 
the mean-field limit. Therefore, in the Fermi case at low temperature, the
low density 
expansion is not adapted to describe the mean-field limit.
 On the other hand, at high density $\lambda^2\rho\gg 1$, (\ref{D2}) leads to
\be\label{Z2} Z={1\over\lambda^2}(1-{1\over2}\lambda^2\rho\alpha+
{1\over2}(\lambda^2\rho)^2\alpha^2[{1\over6}+{1\over\lambda^2\rho}
(\sqrt{\pi
\over\lambda^2\rho }-{1\over\lambda^2\rho}+{1\over\lambda^2\rho}
e^{-\lambda^2\rho}
O({1\over\lambda^2\rho}))]+\ldots \ee
where the leading mean magnetic field term is indeed present.
\hfill\break
\hfill\break
{\bf 3. Fermion case at zero Temperature: the Landau regime}
\hfill\break
\hfill\break
{\it 3.1. The ordered phase}

Before the system reaches the mean-field limit (\ref{mean}),
we expect \cite{mdis} an intermediate
regime caracterised by smooth Landau oscillations in the spectrum.
This intermediate regime is identified  as the ordered phase by opposition
to the
phase with  no oscillation (disordered phase).
The order $\alpha^2$ is quite
instructive to give information on the way the system reaches
the mean-field limit.
Consider indeed the case of Poissonian impurities (\ref{bibi}).
We observe $1/\rho$ corrections to the mean-field
at any order of the perturbative expansion in $(\alpha\rho)^n$.
On the other hand, (\ref{Z2})  shows corrections to the mean-field of 
order $1/(\rho\sqrt\rho)$. This
implies that the system approaches
more rapidly its  mean-field limit
when the impurities are fermions at  zero temperature,  rather than Poissonian.
In other words, the system is less disordered,
since a Fermi distribution of impurities
is more homogeneous than a  Poissonian one.

Let us generalise these considerations at any order $\alpha^n$ of  perturbative
theory. 
One has to evaluate
$<\rho({\bf r}_1)\rho({\bf r}_2)...\rho({\bf r}_n)>$, which can be rewritten
as
\be\label{corr}
<\rho({\bf r}_1)..\rho({\bf r}_n)>=\sum_{p=1}^n\sum_{f\in S_n^p}
{1\over p!}\int d{\bf r}'_1..d{\bf r}'_p\rho({\bf r}'_1,..,{\bf r}'_p)
\prod_{q=1}^n\delta({\bf r}_q-{\bf r}'_{f(q)})\ee
$S_n^p$ is the set of all possible surjections from $(1,..,n)$ to $(1,..,p)$
and $\rho({\bf r}_1,..,{\bf r}_p)$ is the $p$-body correlation function
\be \rho({\bf r}_1,..,{\bf r}_p)=\sum_{\sigma\in S_p}\epsilon(\sigma)
g_b({\bf r}_1-{\bf r}_{\sigma(1)})...g_b({\bf r}_p-{\bf r}_{\sigma(p)})\ee
In the case of Fermions at zero temperature, one has the correlator
\be g_b({\bf r})={1\over r}\sqrt{\rho\over\pi}J_1(\sqrt{4\pi\rho}r)\ee
For example, using
\be <\rho({\bf r_1})\rho({\bf r_2})>=\rho^2-{1\over\vert{\bf r}_1-{\bf r}_2
\vert^2}{\rho\over\pi}[J_1(\sqrt{4\pi\rho}\vert{\bf r}_1-{\bf r}_2\vert)]^2
+\rho\delta({\bf r}_1-{\bf r}_2)\ee
altogether with (\ref{devel}), yields the contibution 
(\ref{D2}),  in addition to the mean-field term.

At high density ($\lambda^2\rho\gg1$), because of  Fermi exclusion,
 the $p$-body correlation function becomes
\be\label{tata} \rho({\bf r}_1,..,{\bf r}_p)=\rho^p[1-{1\over\rho}\sum_{i<j}
\delta({\bf r}_i-{\bf r}_j)+O({1\over\rho\sqrt\rho})]\ee
When  (\ref{tata}) is used for evaluating
$<\rho({\bf r}_1)\rho({\bf r}_2)...\rho({\bf r}_n)>$,
one indeed finds that corrections
to the mean-field term $(\rho\alpha)^n$ are of order $1/(\rho\sqrt\rho)$.
\hfill\break
\hfill\break    
{\it 3.2. Absence of a pure disordered phase}
\hfill\break

We have just seen that  corrections to the average magnetic field limit 
are less important in the Fermi case. Could it be that 
the statistics of the impurities alter
the occurence of the transition itself \cite{mdis}? 
First let us remark that the
average partition function can be expressed as 
\be \label{Zsc}Z={1\over\lambda^2}F(\lambda^2\rho,\lambda^2_v\rho,\alpha)\ee
which means that the average density of states is a function of $E/\rho$,
$\lambda^2_v\rho$ and $\alpha$.
This is due to the fact that $g_b({\bf r})$ is $\rho$ times a function
of $\sqrt\rho r$ and $\lambda^2_v\rho$.
In (\ref{devel}) together with (\ref{corr}) , rescaling $\beta_i$ into
$\beta_i/\beta$, ${\bf r}_i$ into ${\bf r}_i/\lambda$ and ${\bf r}'_i$
into $\sqrt\rho{\bf r}'_i$, immediately leads to (\ref{Zsc}).
In particular at $T_v=0$, $\lambda^2 Z$ is a function of $\lambda^2\rho$ and
$\alpha$.
Considering the expansion
\be \lambda^2 Z=1+{1\over2}\alpha(\alpha-1)\lambda^2\rho+
c_2(\alpha)(\lambda^2\rho)^2+c_3(\alpha)(\lambda^2\rho)^3+\ldots \ee
we can show that $c_2=0$. Use simply 
\be g_b({\bf r})=\rho\sum_{k=0}^\infty{(-\pi\rho r^2)^k\over k!(k+1)!}\ee
and conclude in general that $\rho({\bf r}_1,..,{\bf r}_p)$
is at least of order $\rho^p$. In particular
\be \rho({\bf r}_1,{\bf r}_2)=\rho^2-[g_b({\bf r}_1-{\bf r}_2)]^2\ee
starts at order $\rho^3$, whereas $\rho({\bf r})=\rho$. Therefore,  no
contribution of order $\rho^2$ is to be found in
$<\rho({\bf r}_1)\rho({\bf r}_2)...\rho({\bf r}_n)>$, which means that
the average
partition function does not contain terms of order $\rho^2$ at any order
$\alpha^n$.

The specific heat \cite{mdis}
\be C=k\beta^2{d^2\over d\beta^2}\ln Z\ee
gives a lower bound for the critical value $\alpha_c$ at which the system
does not exhibit any Landau oscillations. More precisely,
if the correction to $C-C_0$ ($C_0=k$) is negative when $\beta\to0$, 
then oscillations
are already present since, at small $\beta$, 
\be C-C_0=2\pi^2k\beta^2\int_0^\infty\int_0^\infty dEdE'{d<\rho(E)/V>\over dE}
{d<\rho(E')/V>\over dE'}(E-E')^2+\ldots \ee
Here, the small  $\beta$ expansion (\ref{Zlow}) yields
\be C-C_0=-{1\over4}k({2\pi\beta\rho\over m})^2\alpha^2(1-\alpha)^2+\ldots \ee
which is always negative.
Therefore the average density of states always displays Landau
oscillations.

\hfill\break
{\bf 4. Conclusion}
\hfill\break

To complete this analysis, let us emphasize again that for
intermediate magnetic impurity temperature, the average partition function
has been shown in (\ref{Zsc}) to scale as  $1/\lambda^2$ times a function
of $\lambda^2\rho$,
$\lambda_v^2\rho$ and $\alpha$. This implies that the average
density of states is a function $<\rho(E/\rho,\lambda_v^2\rho,\alpha)>$. Since
 Poisson distribution
is recovered in the Boltzman limit $T_v\to\infty$ (since
then $\rho({\bf r}_1,..,{\bf r}_p)\to\rho^p$), one
interpolates between the
Poisson and the zero temperature Fermi cases simply  by
varying $\lambda_v^2\rho$ from $0$ to $\infty$. When $\lambda_v^2\rho=0$,
there is a
transition
at $\alpha_c\simeq 0.35$, whereas when  $\lambda_v^2\rho=\infty$, we have just
shown that no
transition occurs at all. Therefore, we expect that, for $\lambda_v^2\rho$
sufficiently small, a transition still occurs at a critical value
$\alpha_c'>0.35$, and for $\lambda_v^2\rho$ sufficiently big, no transition
occurs anymore, meaning that the system is always Landau like, in the whole
interval $\alpha\in[0,1/2]$ 
(Fig 2).
It is not clear  if the transition observed in the Poissonian case
can be interpreted as a phase transition, possibly related 
to a localisation-delocalisation transition. But 
 magnetic impurity distributions do influence this transition 
 by 
actually reordering the system
when the density correlations are increased.
\hfill\break

Acknowledgements : We dedicate this work to the memory of Claude Itzykson,
who suggested to one of us (S.O.) to look at the effect of statistics on the
impurity distribution.

\end{document}